\documentclass[letterpaper,twocolumn,showpacs,amsmath,amssymb]{revtex4}

\makeatletter
\def\ignorecitefornumbering#1{%
     \begingroup
         \@fileswfalse
         #1%                     % do \cite comand
    \endgroup
}
\makeatother

\usepackage{graphicx,color}
\usepackage[colorlinks,plainpages=false,linkcolor=black,urlcolor=blue,citecolor=blue,pdfpagemode=UseNone,pdfstartview=FitH]{hyperref}

\begin{document}

%\title{Electronic structure of FeSe compounds: temperature evolution, nematicity, and superconductivity (Review article)}
\title{Metamorphoses of electronic structure of FeSe-based superconductors}

\author{Yu.~V.~Pustovit}
\author{A.~A.~Kordyuk}
\affiliation{Institute of Metal Physics of National Academy of Sciences of Ukraine, 03142 Kyiv, Ukraine}

\begin{abstract}
The electronic structure of FeSe, the simplest iron based superconductor (Fe-SC), conceals a potential of dramatic increase of $T_c$ that realizes under pressure or in a single layer film.  This is also the system where nematicity, the phenomenon of a keen current interest, is most easy to study since it is not accompanied by the antiferomagnetic transition like in all other Fe-SC's. Here we overview recent experimental data on electronic structure of FeSe-based superconductors: isovalently doped crystals, intercalates, and single layer films, trying to clarify its topology and possible relation of this topology to superconductivity. We argue that the marked differences between the experimental and calculated band structures for all FeSe compounds can be described by a hoping selective renormalization model for a spin/orbital correlated state that may naturally explain both the evolution of the band structure with temperature and nematicity.
\end{abstract}

%\preprint{\textit{xxx}}
\maketitle

\tableofcontents

\section{Introduction}

FeSe, the simplest iron based superconductor \cite{Hsu2008, Mizuguchi2010}, through a number of its incarnations reveal several puzzling features which could be key milestones to understanding the high temperature superconductivity. The superconducting transition temperature of FeSe in form of single crystal is dramatically increased from about 9 to 38 K under pressure \cite{Medvedev2009} and by means of intercalation \cite{Guo2010}. The combination of both intercalation and pressure results in re-emerging superconductivity at 48 \cite{Sun2012}. The single layer FeSe films on SrTiO$_3$ (STO) substrate push the $T_c$ to about 65 K \cite{Wang2012, Liu2012} and may be even higher \cite{Ge2015}, opening a new frontier for superconductivity \cite{Bozovic2014}.

FeSe is also a system with intricate evolution of the electronic band structure with temperature. In the first place, it is a nematic transition that is associated with spontaneous breaking of the symmetry between the $x$ and $y$ directions in the Fe-plane, reducing group symmetry of the lattice from tetragonal to orthorhombic. It is called ``nematic" and believed to be a result of intrinsic electronic instability because its effect on electronic properties is much larger than expected based on the structural distortion observed \cite{Tanatar2010, Chu2010}. Also, in all other iron based superconductors the nematic transition is closely followed by the the antiferomagnetic (AFM) one of the same orthorhombic symmetry \cite{Cruz2008, Huang2008, Paglione2010} that gives natural reason to believe in mutual relation of these two phases and that the nematic transition can be caused by spin-fluctuations \cite{FernandesNP2014} that, in turn, gives a solid support for the $s^{\pm}$-pairing model \cite{Hirschfeld2011}.

FeSe is quite different from all other Fe-SC's and is a sort of uncomfortable example for spin-fluctuation theories: (1) The nematic transition, which happens for FeSe crystals at about 90 K \cite{McQueen2009}, is not accompanied by the AFM at all. (2) The Fermi surface topology for some of FeSe incarnations, such as mentioned intercalates and single layer films, can hardly support the $s^{\pm}$-pairing \cite{Nekrasov2016, Sadovskii2016}. Now more and more evidences coming in favor of charge induced nematicity in FeSe \cite{Massat2016}.

\begin{figure}[b]
\begin{center}
\includegraphics[width=0.42\textwidth]{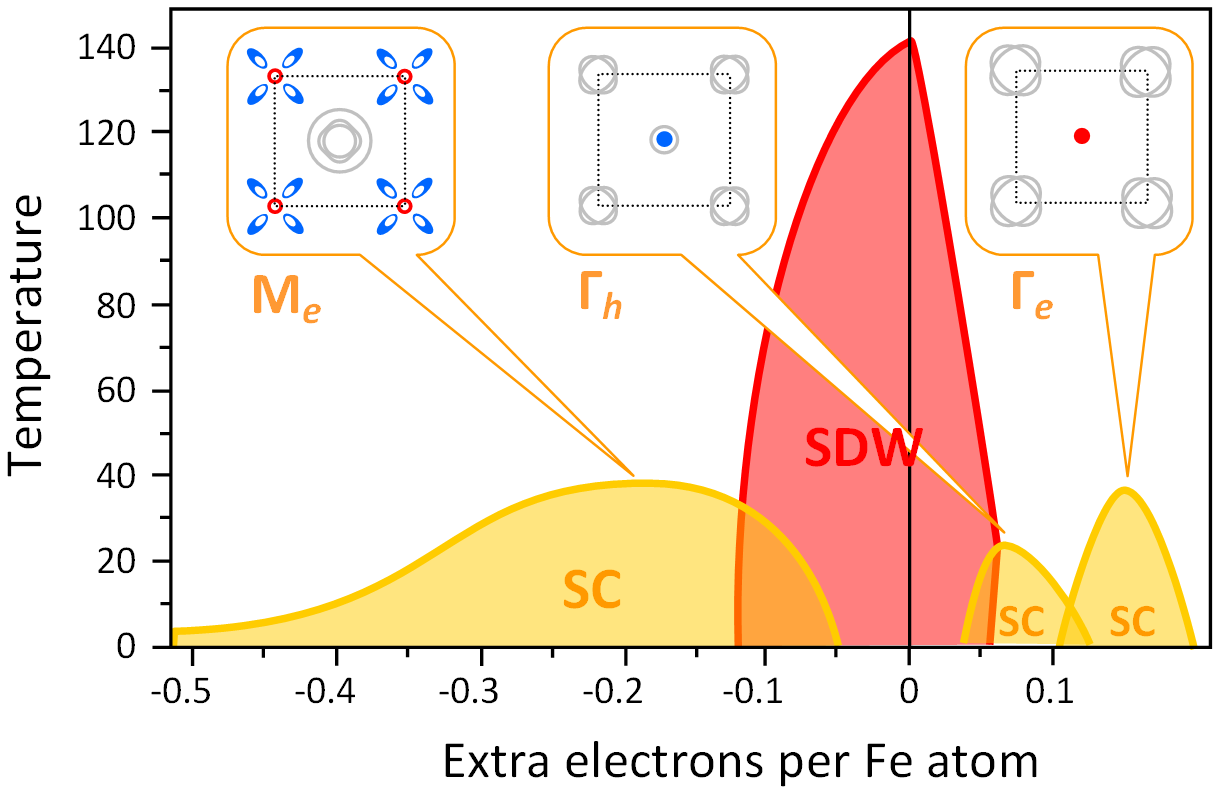}
\caption{Universal electronic phase diagram of Fe-SC's with three superconducting domes that can be classified by a proximity of the corresponding Van Hove singularity to the Fermi level: M$_e$, $\Gamma_h$, and $\Gamma_e$ correspond to proximity to Lifshitz transition of the electron band in M-point and hole and electron bands in $\Gamma$ point, respectively \protect\ignorecitefornumbering{\cite{Kordyuk2012, Kordyuk2013}}.
\label{UPhD}}
\end{center}
\end{figure}

\begin{figure*}
\begin{center}
\includegraphics[width=1\textwidth]{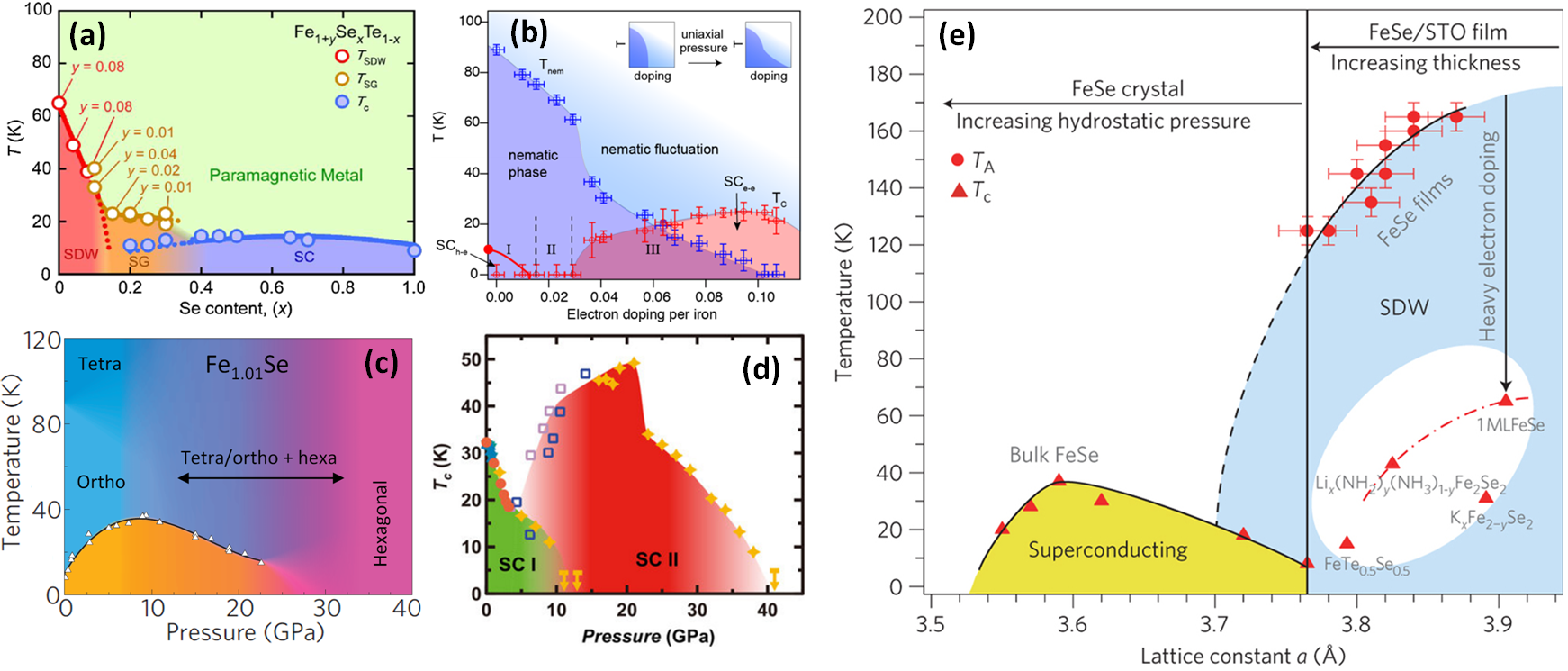}
\caption{Electronic phase diagrams of different FeSe-based compounds: (a) Fe(Se,Te), as an example of isovalent substitution of Te \protect\ignorecitefornumbering{\cite{Katayama2010}}; (b) surface superconductivity on potassium-coated FeSe single crystal \protect\ignorecitefornumbering{\cite{Ye2015}}, (c) FeSe single crystal under pressure \protect\ignorecitefornumbering{\cite{Medvedev2009}}, (d) double-dome superconductivity in ammoniated metal-doped FeSe under pressure \protect\ignorecitefornumbering{\cite{Izumi2015}}, (e) a universal phase diagram for the superconducting and structural transitions vs. lattice constant for all FeSe families \protect\ignorecitefornumbering{\cite{Tan2013}}.
\label{PhDs}}
\end{center}
\end{figure*}

It is also interesting to see whether the FeSe compounds follow the general correlation that $T_c$ is maximal when a certain Van Hove singularity crosses the Fermi level \cite{Kordyuk2012, Kordyuk2013}. This correlation can be explained as a shape-resonance-enhanced superconductivity when the shape of the Fermi surface is critically close to the topological Lifshitz transition \cite{Bianconi2013}. In most cases, such Fermi surface criticality can be easily reveled by the angle resolved photoemission (ARPES) \cite{Kordyuk2014}, like for almost all the Fe-SC's, as shown at a universal phase diagram in Fig.\,\ref{UPhD}. In some other cases, like for the hole doped cuprates \cite{Kordyuk2015}, the Fermi surface criticality is hard to resolve in direct experiment. The later can be also the case for some of FeSe compounds.

In any case, the exact knowledge of the electronic structure of FeSe-based compounds should be important for understanding their intriguing physics. In this review we summarize the results on electronic band structure of different FeSe incarnations, comparing the results of band structure calculations to ARPES experiment, and examine its evolution with temperature, discussing its possible reasons and consequences.

\begin{figure*}
\begin{center}
\includegraphics[width=1\textwidth]{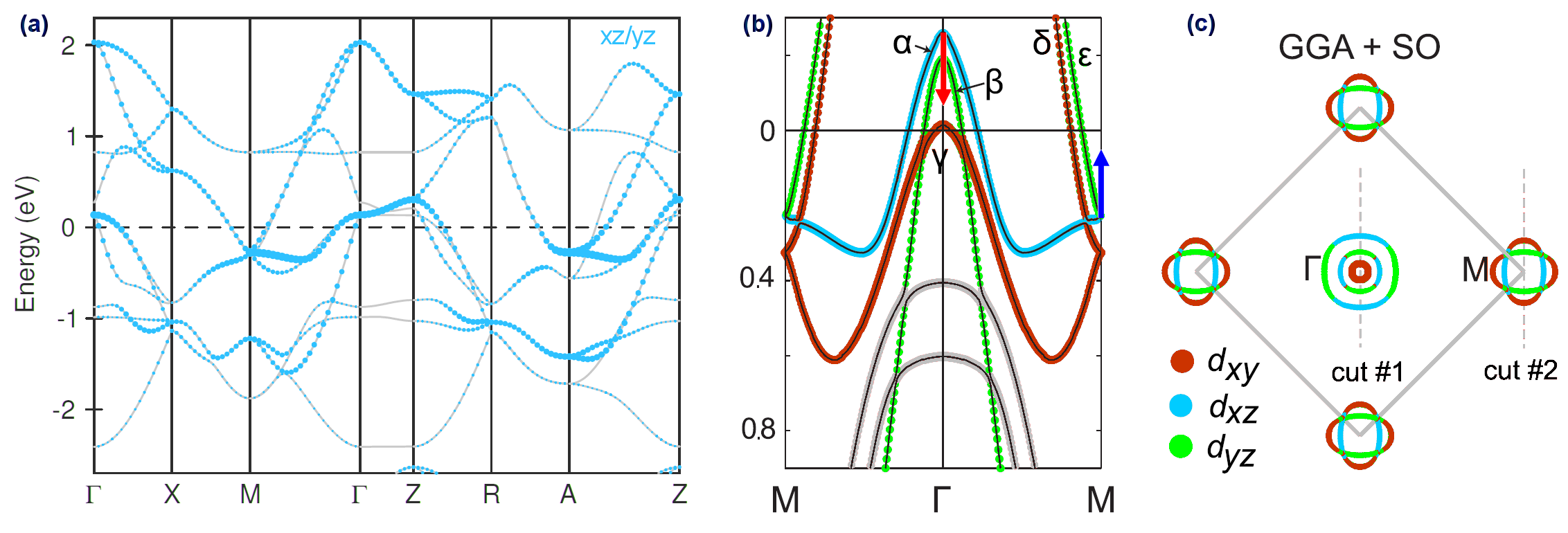}
\includegraphics[width=1\textwidth]{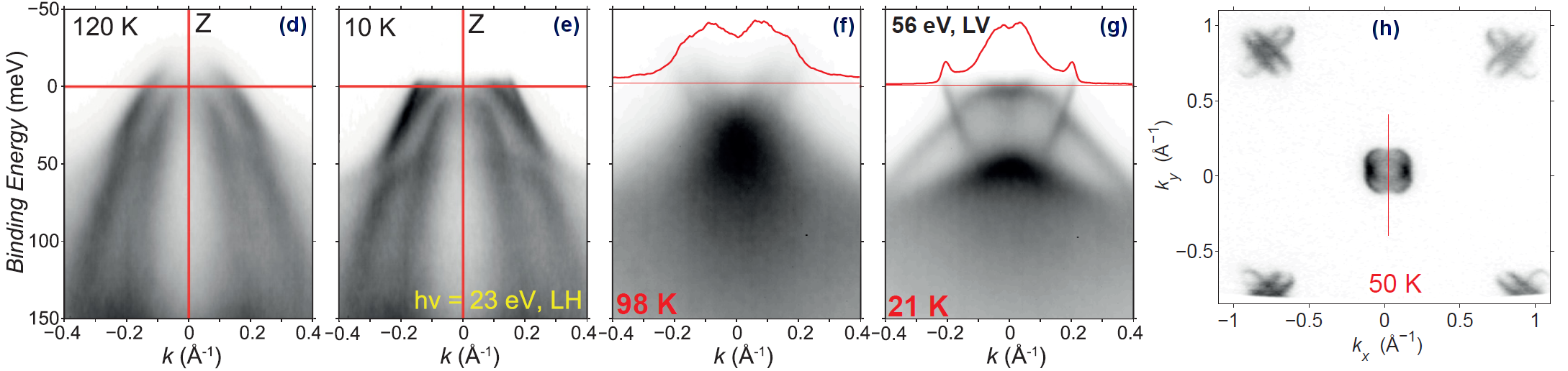}
\caption{Electronic structure of FeSe: (a) non-spin-polarized LDA-LMTO calculations of FeSe band structure, including $xz/yz$ orbital character \protect\ignorecitefornumbering{\cite{Fedorov2016, Maletz2014}};
(b, c) spin-orbit coupling (SOC) included calculation of the band structure and Fermi surface of FeSe in the tetragonal phase, projected into the $k_z = 0$ plane and colored by the dominant orbital character \protect\ignorecitefornumbering{\cite{Watson2015}}; ARPES images of the Brilouine zone (BZ) center close to (0, 0, $\pi/c$) or Z-point ($h\nu$ = 27 eV) (d, e) and close to ($\pi/a$, $\pi/a$, $\pi/c$) or A-point ($h\nu$ = 56 eV) (f, g) above (d, f) and below (e, g) $T_s$; (h) the Fermi surface map measured at 50 K with 56 eV photons \protect\ignorecitefornumbering{\cite{Watson2015, Watson2016}}.
\label{FeSe_elstr}}
\end{center}
\end{figure*}

\begin{figure*}
\begin{center}
\includegraphics[width=1\textwidth]{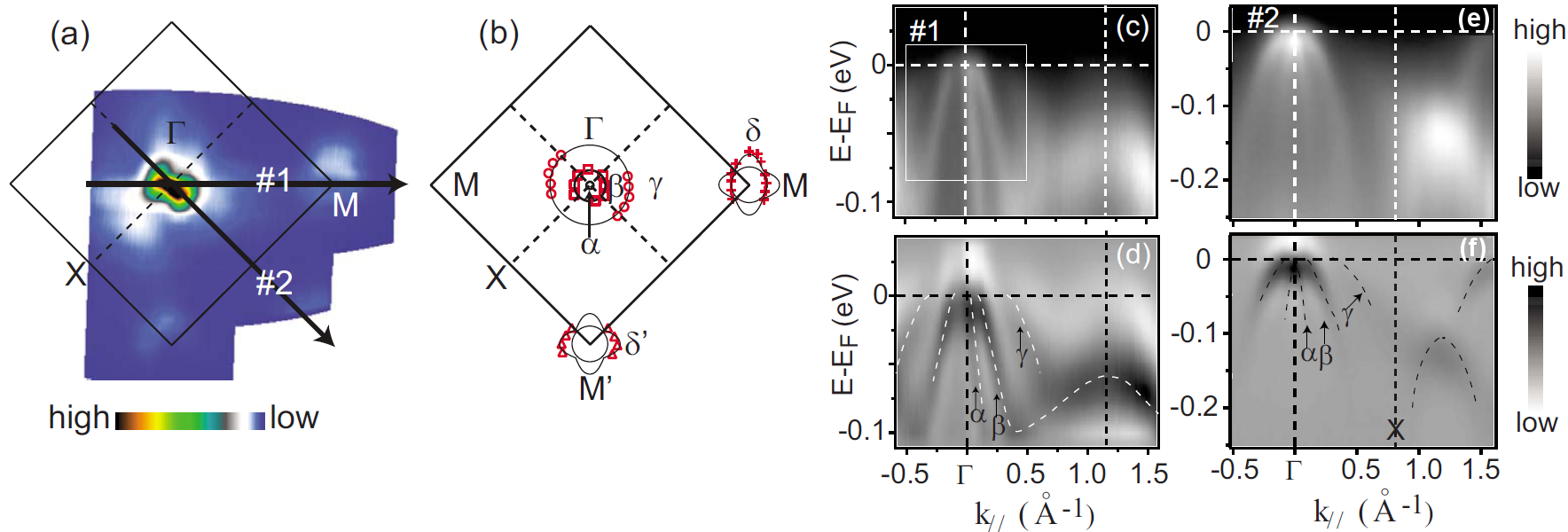}
\caption{Electronic structure of Fe$_{1.04}$Te$_{0.66}$Se$_{0.34}$: (a) measured and (b) reconstructed Fermi surfaces, and (c-f) ARPES images measured with circularly polarized 22 eV photons along the cut \#1 in the $\Gamma$-M direction (c), and along the cut \#2 in the $\Gamma$-X direction (e), as well as their second derivatives with respect to energy (d, f) \protect\ignorecitefornumbering{\cite{Chen2010}}.
\label{FeSeTe_elstr}}
\end{center}
\end{figure*}

The paper is organized as following. In the main Sec.\;\ref{incarnations} we consider three FeSe incarnations, starting from their phase diagrams (Fig.\,\ref{PhDs}) and experimental manifestations of the phase transitions, mainly in transport measurements (Fig.\,\ref{Res}). Then we show the examples of calculated and measured electronic band structure for: single crystals of FeSe (Fig.\,\ref{FeSe_elstr}) and Fe(Se,Te) (Fig.\,\ref{FeSeTe_elstr}) in Sec.\;\ref{crystals}; intercalates (Fig.\,\ref{KFeSe_elstr}) in Sec.\;\ref{intercalates}, where we also discuss the orbital selective renormalization (Fig.\,\ref{Renorma}); and one unite cell films (Fig.\,\ref{1UC_elstr}) in Sec.\;\ref{1UC}. Finally, in Sec.\;\ref{elstr} we summarize the electronic band structure of those three families in Fig.\,\ref{Bands} and Tables\;\ref{tab1} and  \ref{tab2}. Then we briefly discuss the issues of nematicity and evolution of the electronic structure with temperature above the nematic transition in Sec.\;\ref{Nematicity} (Fig.\,\ref{fignemat}).

\section{F\lowercase{e}S\lowercase{e} incarnations}
\label{incarnations}

\begin{figure*}
\begin{center}
\includegraphics[width=1\textwidth]{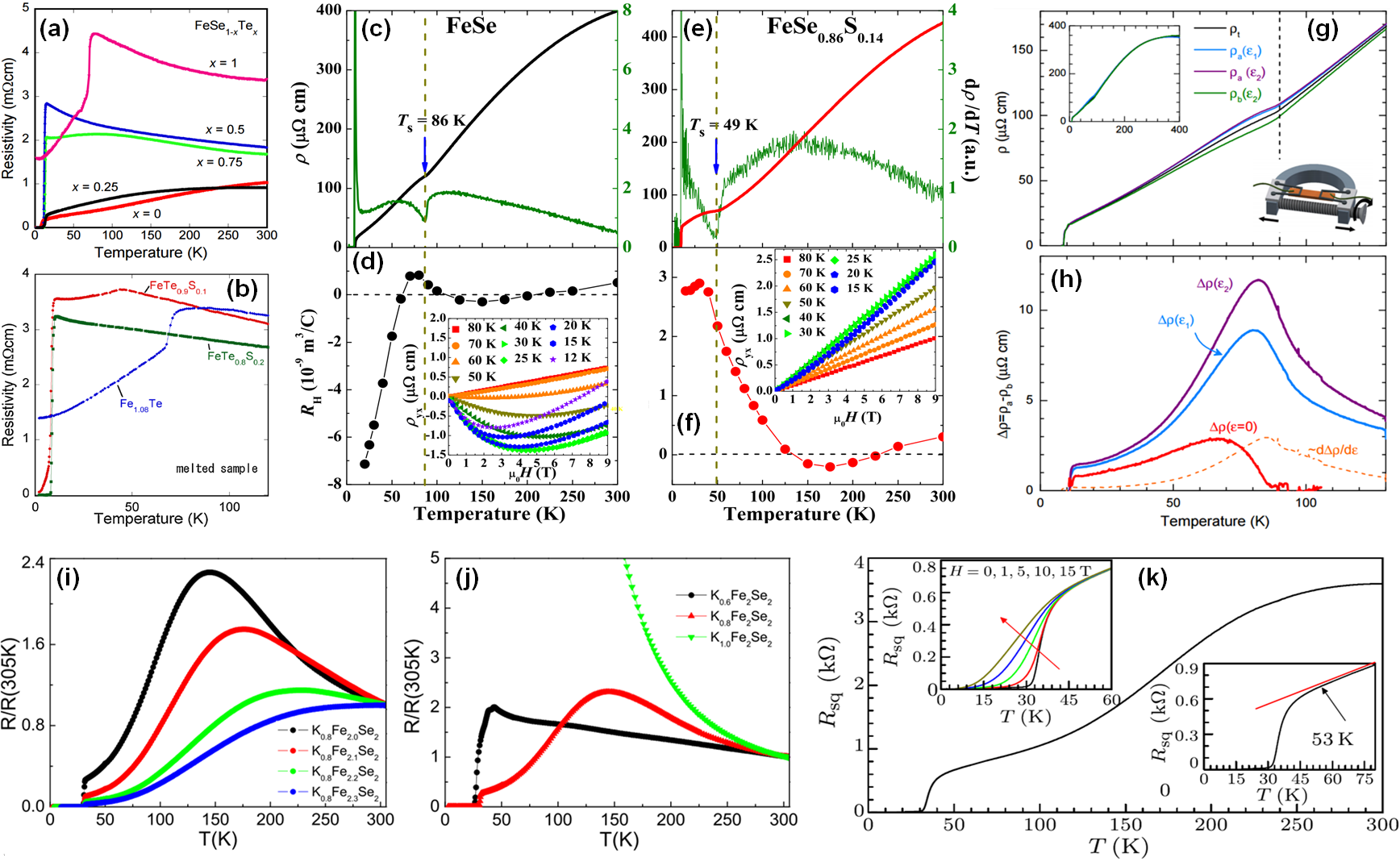}
\caption{Electronic transport in FeSe-based compounds. Resistivity of isovalently-substituted FeSe$_{1-x}$Te$_x$ (a) and FeTe$_{1-x}$S$_x$ (b) \protect\ignorecitefornumbering{\cite{Mizuguchi2010}} show three types of phase transions: antiferromagnetic (AFM), $T_a$, superconducting, $T_c$, and structural, $T_s$, also called ``nematic". The latter is better seen on resistivity (c,\,e) and Hall coefficient (d,\,f) for pure FeSe (c,\,d) and FeSe$_{0.86}$S$_{0.14}$ (e,\,f) \protect\ignorecitefornumbering{\cite{Sun2016}}. The in-plane resistivity anisotropy in strain-detwinned single crystals of FeSe (g) and elastoresistivity measurements allow to extract the intrinsic resistivity anisotropy of strain-free samples (h) that peaks slightly below $T_s$ \protect\ignorecitefornumbering{\cite{Tanatar2015}}. The normal state resistivity of intercalated K$_x$Fe$_{2+y}$Se$_2$ changes dramatically with both K (i) and Fe (j) slight variations \protect\ignorecitefornumbering{\cite{WangDM2011}}. Square resistivity of a 5-UC-thick FeSe film on insulating STO(001) surface (k) \protect\ignorecitefornumbering{\cite{Wang2012}}.
\label{Res}}
\end{center}
\end{figure*}

\textbf{Phase diagrams.} Some examples of the phase diagrams of different FeSe-based compounds are shown in Fig.\,\ref{PhDs}. Evidently, pure FeSe crystal is not optimal for superconductivity since the transition temperature increases with isovalent doping (a) \cite{Katayama2010, Mizuguchi2010, Fedorchenko2011}, surface doping (b) \cite{Ye2015}, and pressure (c) \cite{Medvedev2009, Jung2015}. At the same time, the nematic phase is suppressed with doping and pressure and seems to be competing to superconductivity.

There are many possible selenium-tellurium-sulfur combinations to study the isovalent doping in Fe(Se,Te,S) \cite{Mizuguchi2010}. The most studied ternary system is FeSe$_{1-x}$Te$_x$ \cite{Katayama2010}, though its phase diagram is still not known for the whole doping range \cite{Mizuguchi2010}, although the quality of the crystals is constantly improving \cite{Chareev2013}. Except still missing regions, there is a region in between AFM and superconducting phases that has been considered as a weak superconductivity \cite{Deguchi2012}. The width of this region depends on sample treatment: from AFM to $x = 0.5$ for as-grown samples, but decreases considerably after annealing in oxygen. The transition temperature is also slightly increasing with sulphur doping Fe(Se,S) \cite{AbdelHafiez2016} and emerges in Fe(Te,S) from non-superconducting FeTe and FeS \cite{Mizuguchi2010}, though it has been shown recently \cite{Lai2015} that the later starts to superconduct below 5 K.

\textbf{Phase transitions.} The resistivity curves in isovalently-substituted FeSe$_{1-x}$Te$_x$ (a) and FeTe$_{1-x}$S$_x$ \cite{Mizuguchi2010}, shown in Fig.\,\ref{Res} (a,\,b), demonstrate three types of phase transions: antiferromagnetic (AFM), $T_a$, superconducting, $T_c$, and structural, $T_s$, also called ``nematic". The latter does not depend on magnetic field \cite{Roessler2015} and is well seen on resistivity (c,\,e) and Hall coefficient (d,\,f) for the pure FeSe (c,\,d) and FeSe$_{0.86}$S$_{0.14}$ (e,\,f) \cite{Sun2016}. The in-plane resistivity anisotropy in strain-detwinned single crystals of FeSe (g) and elastoresistivity measurements allow to extract the intrinsic resistivity anisotropy of strain-free samples (h) that peaks slightly below $T_s$ \cite{Tanatar2015}.

The superconducting gap values in FeSe differ essentially from one experiment to another. For example, the gaps determined from ARPES are equal to 1.5 and 1.2 meV in the center and corner of the Brillouin zone (BZ), respectively \cite{Borisenko2016}. The tunneling spectroscopy usually gives larger value of 2.2 meV \cite{Song2011}. In nonlinear conductivity of point contacts, two gaps with 2.5 and 3.5 meV were identified \cite{Naidyuk2016}. This discrepancy may result from inhomogeneity or complexity of the band structure: different probes can be more sensitive to different bands with different gap values, not to say affected by close vicinity of several van Hove singularities (VHs's) to the Fermi level.

In contrast to superconducting ferro-pnictides, the layers of iron chalcogenides are neutral and kept together by weak van der Waals interaction. Therefore, intercalation by atoms and molecules is the most easy way to modify their structure (for recent review see \cite{Vivanco2016}). The first intercalated FeSe compounds A$_{x}$Fe$_{2-y}$Se$_{2}$ (A = K,\,Rb,\,Cs) have shown $T_c$ up to 30K \cite{Guo2010} but it has not been straightforward to determine the structure of the superconducting phase \cite{Maletz2013}. From Fig.\,\ref{Res} (i,\,j) one can see that the normal state resistivity of intercalated K$_x$Fe$_{2+y}$Se$_2$ changes dramatically with both K (i) and Fe (j) slight variations \cite{WangDM2011} and it has been shown that the superconducting phase is sandwiched between two AFM insulating phases on the electronic phase diagram as a function of Fe valence \cite{Yan2011}.

The structures of the intercalates like K$_{0.8}$Fe$_{1.7}$Se$_{2}$ or Tl$_{0.6}$Rb$_{0.4}$Fe$_{1.67}$Se$_{2}$ may be optimal for superconductivity since $T_c$ = 32 K seems to be maximal but starts to decrease with pressure  \cite{Sun2012}. On the other hand, $T_c$ = 30 K does not look maximal for the structure of the ammoniated metal-doped iron selenide, (NH$_3$)$_y$Cs$_{0.4}$FeSe, since superconductivity starts to decrees rapidly with pressure \cite{Izumi2015}, see Fig.\,\ref{PhDs}\,(d). But in both cases, a superconducting phase with much higher $T_c$ up to 49 K appears at higher pressure: at 12 and 21 GPa, respectively \cite{Sun2012, Izumi2015}.

One unit-cell (1UC) thick FeSe films grown on a Se-etched SrTiO$_3$ (STO) (001) substrate have shown the superconducting gap about 20 meV in tunneling spectra \cite{Wang2012}. Based on this value, it was concluded that $T_c$ could be about 80 K, assuming the same superconducting mechanism as for the bulk FeSe with $2\Delta/k_B T_c \approx 5.5$. The resistivity of 1UC-thick film is tricky to measure and in Ref.\,\onlinecite{Wang2012} only the resistivity of 5-UC-thick FeSe film, shown in Fig.\,\ref{Res}\,(k), has been presented. The superconducting transition happens above 30 K, and the authors could even say that it starts above 50 K. The later measurements however have shown that $T_c$ is really high and the todays record is slightly above 100 K by in situ four-point probe electrical transport measurements \cite{Ge2015}. There is also a number of ARPES data that show the superconducting gap closing above 60 K \cite{Liu2012, He2013, Tan2013}, but those ARPES results we will discuss in more details in the sections below.

A universal phase diagram for the superconducting and structural transitions vs.\,lattice constant for all mentined FeSe families \cite{Tan2013} is shown in Fig.\,\ref{PhDs}\,(e). Together with the two-dome phase diagrams of intercalates on pressure \cite{Sun2012, Izumi2015}, like shown in Fig.\,\ref{PhDs}\,(d), it may suggest either two different mechanisms of pairing or two different peculiarities of the electronic band structure responsible for superconductivity. The latter possibility we discuss below, but before should briefly address the issue of the spin ordering and its possible relation to nematicity.

\textbf{Magnetic ordering.} First, unlike the Fe-pnictides \cite{Chubukov2015}, the spin-driven nematic scenario for FeSe crystals has been considered as unlikely based on thermal-expansion \cite{Bohmer2013} and NMR data \cite{Baek2015, Bohmer2015}. Ab initio calculations indicated that FeSe is close to magnetic instability \cite{Grechnev2012} but no magnetic order has been observed in FeSe thus far \cite{McQueen2009, Roessler2016}, that could be explained by strong frustration of the magnetic fluctuations \cite{Glasbrenner2015} or by formation of a quantum paramagnet \cite{Wang2015}. Only spin fluctuations around the AFM wave vector were found in Fe(Te,Se) by inelastic neutron scattering \cite{Li2009}. Other neutron scattering experiments show that these spin fluctuations are coupled with orthorhombicity \cite{Rahn2015, Wang2016}. On the other hand, it has been suggested that the smallness of the Fermi energy in FeSe leads to a near-degeneracy between magnetic fluctuations and fluctuations in the charge-current density-wave channel \cite{Chubukov2015}: if $E_F \sim T_s$, both a spin density wave (SDW) channel and a charge-current density-wave (iCDW) channel are comparable and strongly fluctuating at the nesting vector. One may conclude that in order to resolve the spin- vs orbital fluctuations dilemma, the exact knowledge of the electronic structure with its orbital origin is required \cite{Kordyuk2012, Evtushinsky2014PRB}.

\subsection{Single crystals}
\label{crystals}

The electronic structures of FeSe single crystal, calculated \cite{Maletz2014, Watson2015, Fedorov2016} and measured by ARPES \cite{Watson2015, Watson2016}, are presented in Fig.\,\ref{FeSe_elstr}. Like in all other Fe-SC's, the metallic properties are defined by Fe $3d$ bands and, in calculated band structure (see also \cite{Subedi2008, Aichhorn2010}), there are five bands crossing the Fermi level that form five Fermi surface sheets: three around the center of the BZ ($\Gamma$-point), and two around its corners (M-point). These bands are formed mainly by $d_{xy}$, $d_{xz}$, and $d_{yz}$ orbitals, as shown on the upper panels. The most representative is the cut in the 2Fe BZ taken along $\Gamma$M-direction. Along this direction the main orbital character for each band remains the same, if the spin-orbit coupling (SOC) is not taken into account: $d_{xz}$ for $\alpha$ band, $d_{yz}$ for $\beta$ and $\varepsilon$ bands, and $d_{xy}$ for $\gamma$ and $\delta$ bands. SOC results in hybridization (splitting $\sim$ 20 meV) between $\alpha$ and $\beta$ bands in the BZ center and near their crossings with the $\gamma$-band, (b) \cite{Cvetkovic2013, Maletz2014, Borisenko2016, Watson2015}. There is also essential $k_z$ dispersion that mainly affects the $d_{xz}$ and $d_{yz}$ bands, as one can see in panel (a) \cite{Maletz2014}, where $\Gamma = (0,0,0)$, Z = $(0,0,\pi)$, M = $(\pi,\pi,0)$, A = $(\pi,\pi,\pi)$ in 2Fe BZ. Note that in Z-point the splitting between $\alpha$ and $\beta$ bands is present even without SOC since $\beta$ band there has essential admixture of Se $d_{3z^2-1}$ orbital. Similarly, one can see strong $k_z$ dependence of $\gamma$ and $\delta$ bands along M-A direction due to strong admixture of $d_{xz}/d_{yz}$ orbitals near M-point.

The band structure seen by ARPES differs from the calculated one in mostly the same way as for all other Fe-SC's \cite{Kordyuk2012, Kordyuk2013}: by the overall band renormalization (presumably a result of coupling to electronic excitations, peaked at about 0.5 eV \cite{Evtushinsky2014}) and by shifting of $\Gamma$ and M-band bunches in the opposite directions, as shown in Fig.\,\ref{FeSe_elstr}\,(b) by the red and blue arrows. Peculiar for FeSe is that the renormalization of the $\gamma$ ($d_{xy}$) band is about 5 \cite{Fanfarillo2016} (one of the moderate estimates, comparing to 9 \cite{Maletz2014} or 17 \cite{Tamai2010}, though the latter was given for ${\mathrm{FeSe}}_{0.42}{\mathrm{Te}}_{0.58}$) that is essentially larger than for Ba-122 family \cite{Kordyuk2013}, for example. Also, the $\gamma$ band sinks below the Fermi level at $\Gamma$ and usually is hardly visible for ARPES but it hybridizes with $\alpha$ and $\beta$ bands (that is often the only way it can be detected). The band structure changes dramatically below 90 K, as one can see in Fig.\,\ref{FeSe_elstr} comparing panels (d,\,f) to (e,\,g) \cite{Watson2015, Watson2016}, but we will discuss it later, in Section\,\ref{Nematicity}.

In the first column of Fig.\,\ref{Bands} we summarize the calculated (top) and experimental (bottom) electronic band structure of FeSe single crystals. In the former, the blue and brown dotted lines show the effect of SOC, but here we will refer to the corresponding solid lines ($\alpha$ and $\gamma$), to keep the same orbital origin along each band. The experimental bands were fitted to the spectra (mainly from \cite{Watson2015, Watson2016}) at about 100 K, that is a bit above the nematic transition.

\begin{figure*}
\begin{center}
\includegraphics[width=1\textwidth]{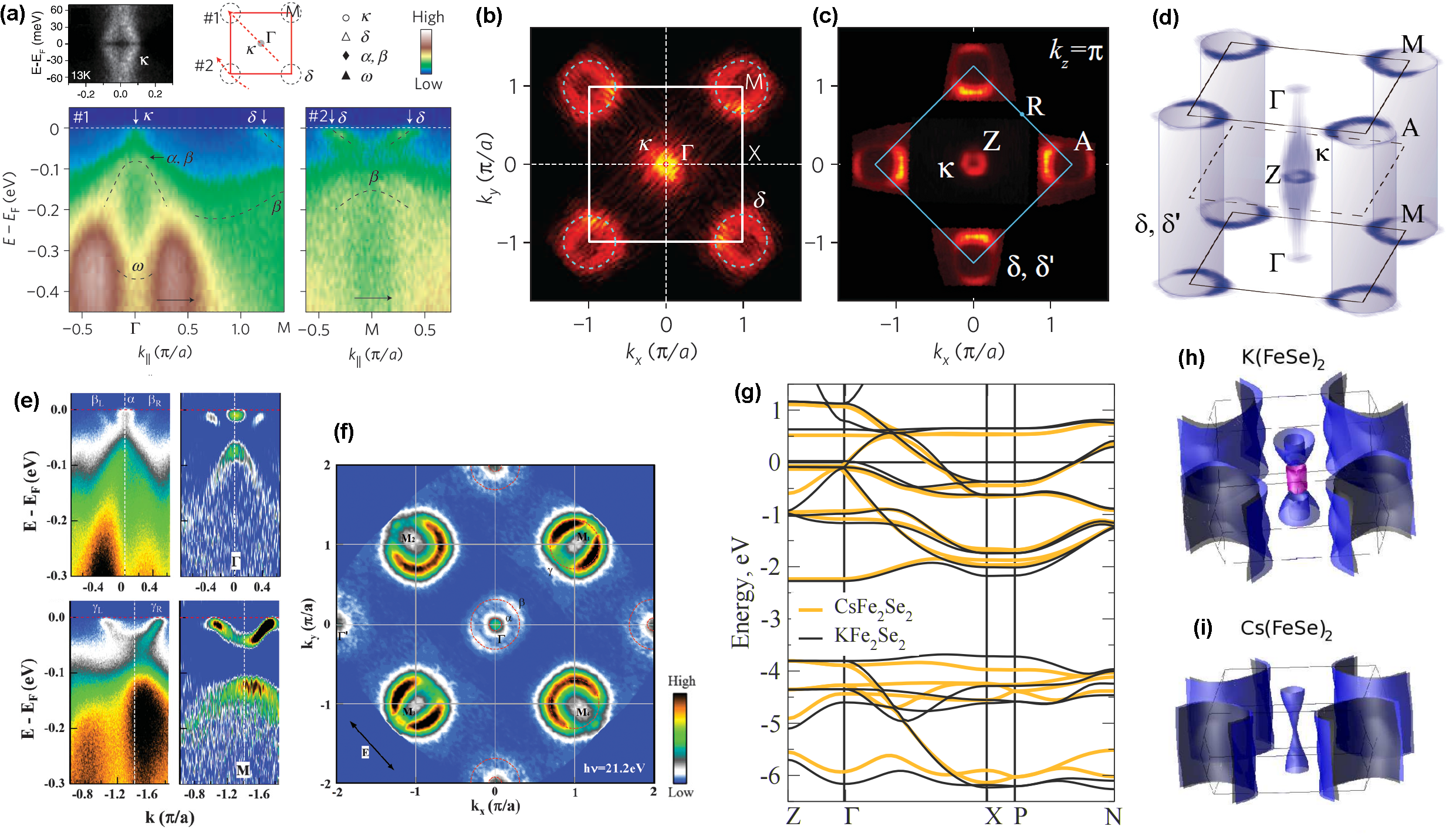}
\caption{Electronic structure of intercalated FeSe:
(a) ARPES spectra of K$_x$Fe$_{2-y}$Se$_2$ (namely K$_{0.8}$Fe$_2$Se$_2$) along cut \#1 or $\Gamma$-M in the Brillouin zone, as shown at the top, and along cut \#2 across the zone corner, and (b, c) the Fermi surfaces measured with 21.2 eV and 31 eV photons at 35 K, and its 3D representation (d) \protect\ignorecitefornumbering{\cite{Zhang2011, Xu2012}}.
(e, f) The photoemission spectra of $({\mathrm{Tl}}_{0.58}{\mathrm{Rb}}_{0.42}){\mathrm{Fe}}_{1.72}{\mathrm{Se}}_{2}$ measured along two high symmetry cuts in $\Gamma$-M direction and corresponding Fermi surface \protect\ignorecitefornumbering{\cite{Mou2011}}.
(g) LDA calculated band structure and (h, i) Fermi surfaces of KFe$_2$Se$_2$ and CsFe$_2$Se$_2$ \protect\ignorecitefornumbering{\cite{Nekrasov2011}}.
\label{KFeSe_elstr}}
\end{center}
\end{figure*}

So, in reality, as a result of the ``red-blue shift" shown in Fig.\,\ref{FeSe_elstr}\,(b), both the hole- and electron-like FS's are essentially smaller than calculated. There is only one hole-like FS formed by the $\alpha$ band. Its evident splitting in two ellipses is attributed to presence of two types of nematic domains \cite{Watson2015, Watson2016}. The $\gamma$ band is 50 meV below $E_F$ at $\Gamma$-point. The $\beta$ band is closer, at about -20 meV in $\Gamma$ and is almost touching the Fermi level in Z-point. Then, from FS topology point of view, FeSe has two VHs's in close vicinity to $E_F$: $\beta$ in $\Gamma$-point and $\varepsilon + \alpha$ in M-point. An explanation why $T_c$ is not very high could be that the former band is still below $E_F$ and rather steep while the later is a result of nematicity that competes with superconductivity in some yet unclear way.

The electronic structure of Fe(Se,Te) differs from FeSe by position of $\gamma$ band, that crosses the Fermi level forming the outer FS around the BZ center, and by position of $\beta$ band that forms small 3D FS \cite{Tamai2010, Chen2010}. The increase of hole FS area should be compensated by larger electron pockets, and rather large shallow pocket around M-point has been observed \cite{Tamai2010}. Fragments of experimental electronic structure of Fe$_{1.04}$Te$_{0.66}$Se$_{0.34}$ are shown in Fig.\,\ref{FeSeTe_elstr} \cite{Chen2010}. So, in terms of empirical correlation between $T_c$ and proximity to Lifshitz transition of $d_{xz}/d_{yz}$ bands \cite{Kordyuk2012}, the electronic structure of Fe(Se,Te) should be more favorable for superconductivity. Alternatively, some enhancement of $T_c$ can be related with the suppressed nematicity.

\subsection{Intercalates}
\label{intercalates}

\begin{figure*}
\begin{center}
\includegraphics[width=1\textwidth]{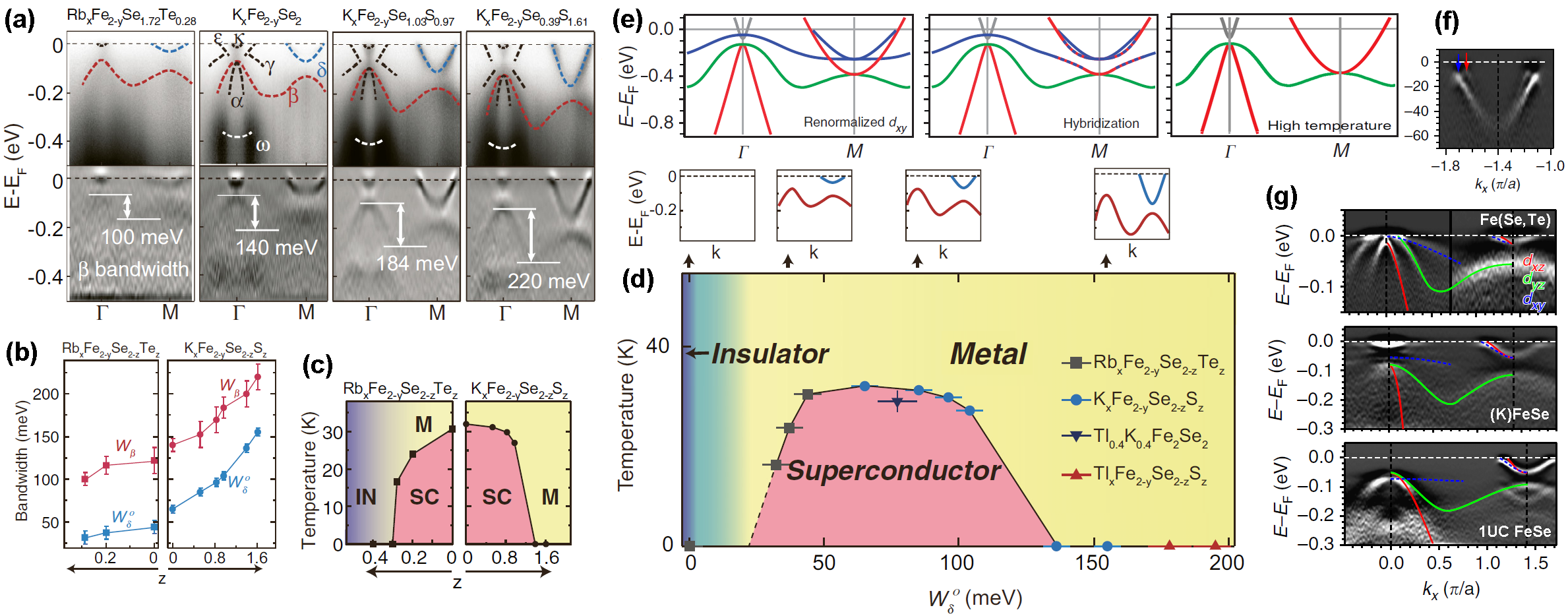}
\caption{Orbital selective strong renormalization: (a-d) correlation-driven superconductor-to-insulator transition for the 122 iron chalcogenides Rb$_x$Fe$_{2-y}$Se$_{2-z}$Te$_{z}$ and K$_{x}$Fe$_{2-y}$Se$_{2-z}$S$_{z}$ \protect\ignorecitefornumbering{\cite{Niu2016}}, and (e-g) universality of strong orbital-dependent renormalization for FeSe-based compounds \protect\ignorecitefornumbering{\cite{Yi2015}}.
(a) Examples from doping dependence of the ARPES spectra along the $\Gamma$-M direction and the corresponding second derivatives with respect to energy (note the different notation of $\alpha$, $\beta$, $\gamma$ bands, comparing to Fig.\,\ref{FeSe_elstr}b);
(b) the bandwidth of the $\beta$-band ($W_{\beta}$) and the occupied width of the $\delta$-band ($W_{\delta}^o$) as a function of doping;
(c) phase diagrams of Rb$_x$Fe$_{2-y}$Se$_{2-z}$Te$_{z}$ and K$_{x}$Fe$_{2-y}$Se$_{2-z}$S$_{z}$ (SC, M, and IN represent the superconducting, metallic, and insulating phases, respectively);
(d) a unified phase diagram of $T_c$ \emph{vs} $W_{\delta}^o$.
(e) Schematics of the effect of orbital-dependent band renormalization, hybridization, and temperature dependence,
(f) high-resolution spectra showing the presence of two electron bands around M, and
(g) rough comparison of the bands along $\Gamma$M direction on top of second energy derivatives of ARPES spectra for FeTe$_{0.56}$Se$_{0.44}$ (top), K$_{0.76}$Fe$_{1.72}$Se$_{2}$ (middle), and 1UC FeSe (bottom).
\label{Renorma}}
\end{center}
\end{figure*}

The intercalated FeSe superconductors are electron doped, so that only the electron-like FS's remain, as one can see in Fig.\,\ref{KFeSe_elstr}. In LDA calculated electronic band structure of KFe$_2$Se$_2$ \cite{Nekrasov2011} the $d_{xy}$ band only touches the Fermi level at $\Gamma$-point. A short recent review of calculations of electronic band structure of FeSe-based superconductors, from intercalates to single layer films is given in \cite{Nekrasov2016}.

The calculated FS is generally supported by a number of published ARPES spectra measured for FeSe intercalated by alkali metals, in particular for A$_x$Fe$_{2-y}$Se$_2$ (A = K, Cs) \cite{Zhang2011, Qian2011, Xu2012}, $({\mathrm{Tl}}_{0.58}{\mathrm{Rb}}_{0.42}){\mathrm{Fe}}_{1.72}{\mathrm{Se}}_{2}$ ($T_c$ = 32 K) \cite{Mou2011}, (Tl,K)Fe$_{1.78}$Se$_{2}$ \cite{Wang2011}, Rb$_{0.77}$Fe$_{1.61}$Se$_{2}$ (32.6 K) \cite{Maletz2013}; or by molecules:  Li$_{x}$(NH$_{2}$)$_{y}$(NH$_{3}$)$_{1-y}$ (43 K) \cite{Burrard-Lucas2013}, Li$_{x}$(C$_{2}$H$_{8}$N$_{2}$)$_{y}$ (45 K) \cite{Hatakeda2013}, (Li$_{0.8}$Fe$_{0.2}$)OH (40 K) \cite{Pachmayr2015, Lu2015, Niu2015}. Evidently, such a FS is away from nesting conditions and its topology does not support the $s\pm$ pairing scenario.

However, there are some differences between calculations and experiment. First, it is a small electron pocket that appears in BZ center: $\kappa$ band in Fig.\,\ref{KFeSe_elstr} (a-c). It makes a small 3D FS centered around Z-point, (d), and according to calculations, consists of Se $d_{3z^2-1}$ and Fe $d_{xz}/d_{yz}$ orbitals. The $\kappa$ pocket is clearly seen in ARPES spectra and accurately studied by many authors \cite{Zhang2011, Xu2012, Maletz2013}.

Second, a larger electron pocket is seen sometimes around $\Gamma$-point \cite{Wang2011, Mou2011} but its origin is not clear. It is not present in LDA calculations and it looks identical to the large pocket around M-point, so, most probably, it is a replica of the pocket from M-point due to superstructure of Se distortions \cite{Zhao2012, Niu2015}. In the papers where only one small $\kappa$ pocket is seen in ARPES spectra \cite{Zhang2011, Xu2012}, two superconducting gaps are found: the smaller one of 7-8 meV opens on this $\kappa$ pocket, and the larger gap of about 10 meV opens on the pocket around M-point. The gap on the extra pocket around $\Gamma$, if observed, is the same as on M-pocket \cite{Wang2011} or even higher \cite{Mou2011}, though it could be because the M-pocket consists of two $\varepsilon$ and $\delta$ bands formed by $d_{xz}/d_{yz}$ and $d_{xy}$ orbitals, respectively, see Fig.\,\ref{Bands}.

Even more complicated picture comes from recent ARPES paper \cite{Sunagawa2016}, where, in addition to two mentioned electron pockets, the hole pocket has been found around the BZ center. It is hardly visible but argued to have mainly $d_{xy}$ origin (invisible part) and partially $d_{xz}$, i.e.\;to to be ``real" $\gamma$ band, according notations from Fig.\,\ref{Bands}. As for the band marked as $\gamma$ in Fig.\,\ref{Bands} (central column), as well as for the aforementioned large central electron pocket, they are supposed to be either from the surface or from another phase \cite{Sunagawa2016}.

The results of LDA+DMFT \cite{Nekrasov2013}, being in general agreement with ARPES data, show that all the Fe $3d$ bands crossing the Fermi level have equal renormalization. On the other hand, detailed study of the isovalently doped A$_x$Fe$_{2-y}$Se(Te,S)$_2$ \cite{Niu2016} shows that the bandwidths of the low energy bands depend on doping. Even when the Fermi surface in the metallic phases is unaffected by the isovalent dopants, the ground state evolves from a metal to a superconductor, and eventually to an insulator when the bandwidth decreases. It has been argued that the band renormalization is strong, orbital-dependent, and universal for all FeSe-based compounds \cite{Yi2015}, see Fig.\,\ref{Renorma}.

Finally, one should note that in the intercalated compounds the intrinsic phase separation is often an issue and the studied crystals are just a mixture of metallic/superconducting and insulating/AFM phases \cite{Maletz2013}.

\subsection{Single layers}
\label{1UC}

\begin{figure*}
\begin{center}
\includegraphics[width=1\textwidth]{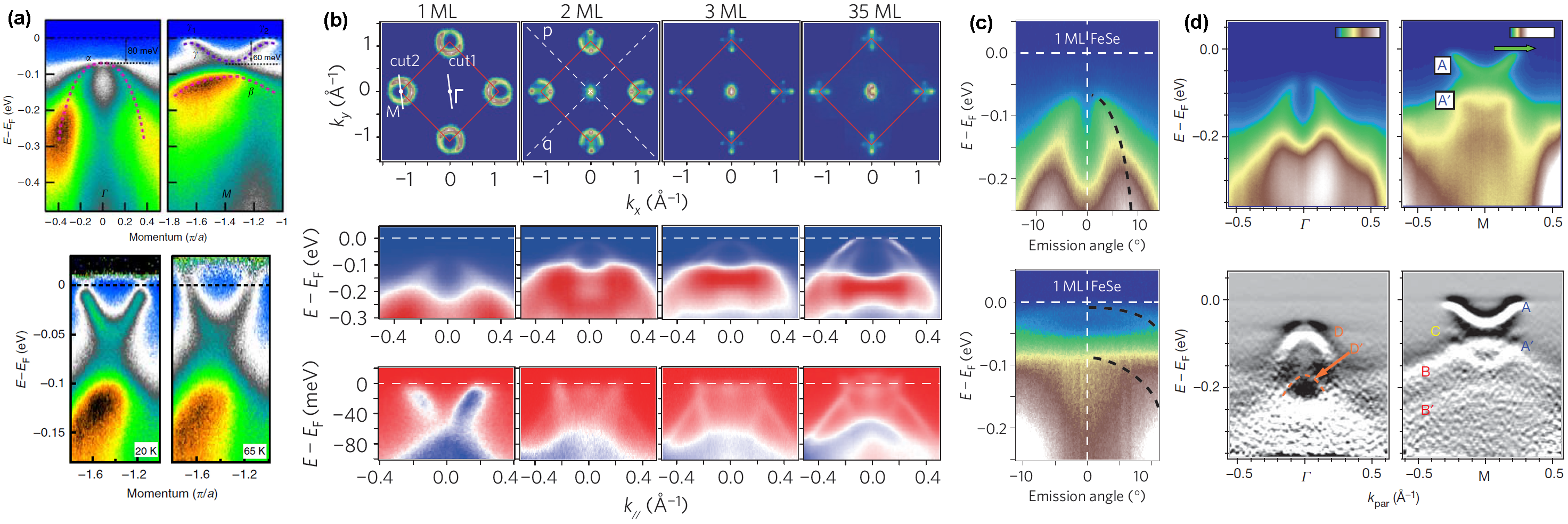}
\includegraphics[width=1\textwidth]{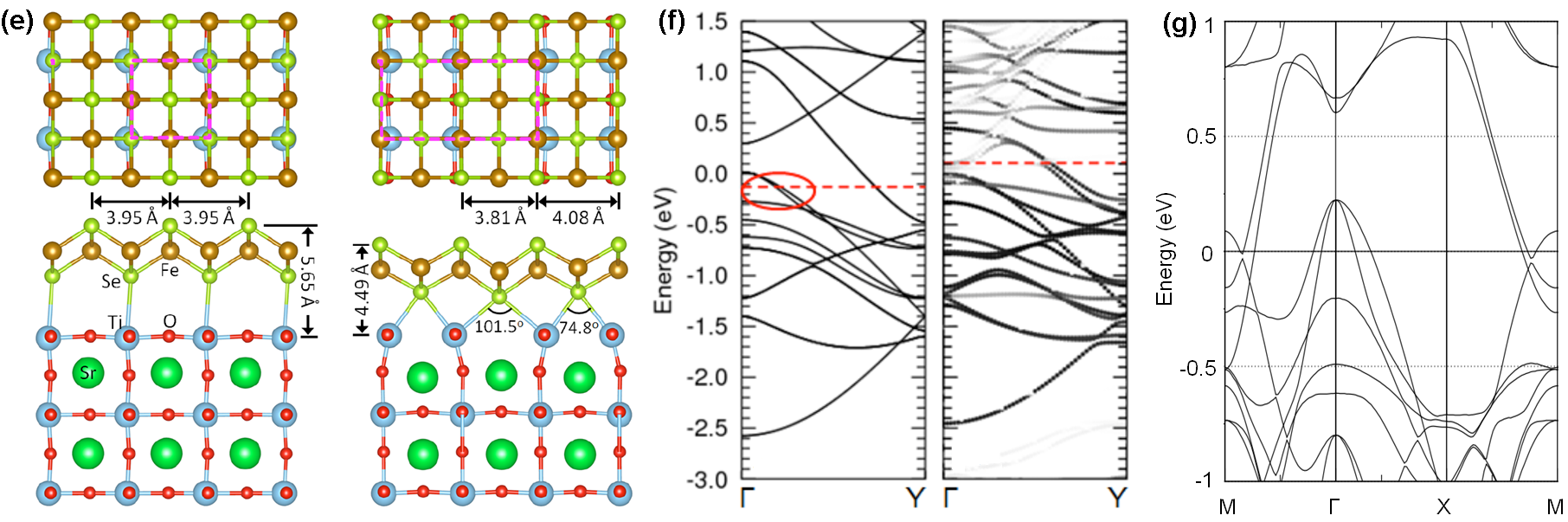}
\caption{Electronic structure of 1UC FeSe on SrTiO$_3$ (STO).
(a) ARPES spectra around $\Gamma$ (top left) and M (top right) points, and temperature evolution of the later (bottom) \protect\ignorecitefornumbering{\cite{Liu2012}}.
(b) Fermi surfaces (top) and ARPES spectra along cut1 (middle) and cut2 (bottom) for 1, 2, 3, and 35 monolayers (ML) \protect\ignorecitefornumbering{\cite{Tan2013}}.
(c) Spectrum around $\Gamma$ of 1UC FeSe measured with 7 eV laser (bottom) and 21.2 eV (top and all other spectra) He lamp \protect\ignorecitefornumbering{\cite{Tan2013}}.
(d, top) High-symmetry cuts measured at 16 K along $\Gamma$M-direction centered at $\Gamma$ (left) and M (right). On right panel a different color scale highlights two important features: the electron band with a minimum at 60 meV below $E_F$ (labeled A), and a replica of electron band (labeled A'), which is located 100 meV below the former and sits on top of a broad hole band. (d, bottom) Second derivatives in energy of the high-symmetry cuts from top row \protect\ignorecitefornumbering{\cite{Lee2014}}.
(e) Structural models of 1UC FeSe on pristine SrTiO3(001) surface (left), which is terminated by a TiO$_2$ layer and on O-deficient surface (right), which is characterized by alternately missing O-atom rows \protect\ignorecitefornumbering{\cite{Bang2013}}.
(f) Band structure of a free-standing 1UC FeSe (left) and deposited on the STO (001) surface containing O vacancies \protect\ignorecitefornumbering{\cite{Bang2013}}.
(g) The LDA calculations of 1UC FeSe on STO substrate that reveal the appearance of additional band of O2$p$ surface states near the Fermi level with good nesting-like matching of the hole Fe3$d$ band \protect\ignorecitefornumbering{\cite{Nekrasov2016}}.
\label{1UC_elstr}}
\end{center}
\end{figure*}

\begin{figure*}
\begin{center}
\includegraphics[width=0.8\textwidth]{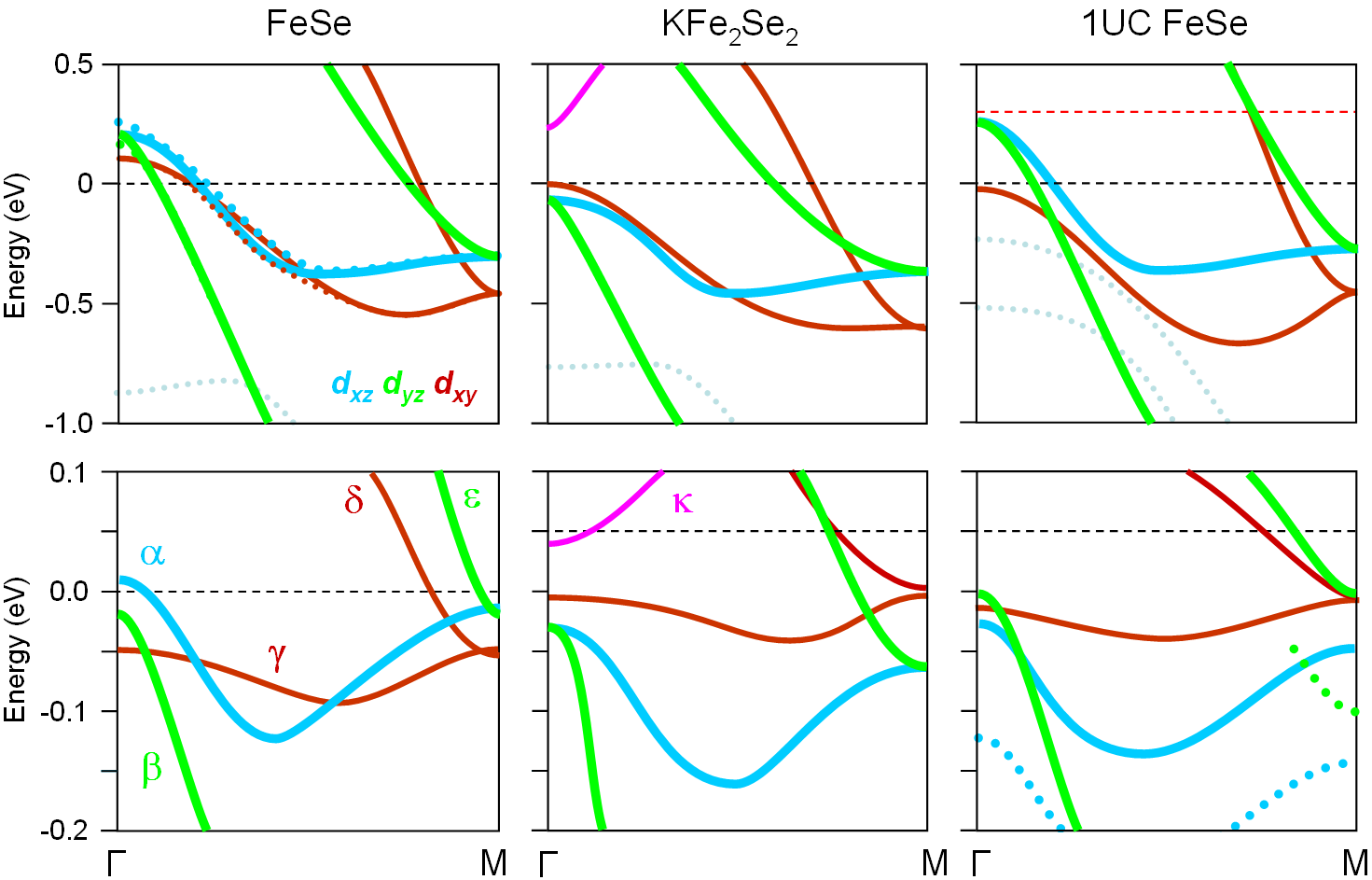}
\caption{Electronic band structure of FeSe-based compounds derived from LDA calculations (top row) and from ARPES experiments (bottom row). The bands are shown in 1.5 eV (LDA) and 0.3 eV (ARPES) energy windows for all the compounds; the Fermi level is marked by the dashed lines in all panels; the upper red dashed line in 1UC FeSe LDA panel corresponds to the electron doping consistent with the Fermi surface area measured by ARPES.
\label{Bands}}
\end{center}
\end{figure*}

The ARPES spectra from the one unit cell (1UC) film on SrTiO$_3$ (STO) substrate have been measured shortly after discovery of high $T_c$ superconductivity in such films \cite{Liu2012, He2013, Tan2013}. The electronic structure consists only of electron-like pockets near the zone corner without indication of any Fermi surface around the zone center, as shown in Fig.\,\ref{1UC_elstr} (a,b) (1ML) \cite{Liu2012}. Panel (d, top) shows the high-symmetry cuts measured at 16 K along $\Gamma$M direction and centered at $\Gamma$ (left) and M (right): the hole band is located 80 meV below $E_F$.

Comparing the intercalates with respect to single crystals was natural to expect that their electronic band structure is (1) more two-dimensional, and (2) shifted below the Fermi level due to electron doping. One can see that the electronic structure of 1UC on STO is remarkably similar to that of intercalates, see Fig.\,\ref{Renorma} (g) or Fig.\,\ref{Bands}, and the superconducting transition temperature is even higher (above 65 K). Since it is the single layer, the band structure is 2D \cite{Nekrasov2016}, by definition, but the reason for electron doping is not quite clear, despite the fact that formation of the electron gas at the interface with the STO is a widely known phenomenon, studied for a long time \cite{Ohtomo2004, SantanderSyro2011}.

First, it was suggested \cite{Wang2012} that STO surface is terminated by a TiO$_2$ layer, but, as shown by \cite{Bang2013}, O-deficient surface better explains the observed electron doped FS \cite{Liu2012, He2013, Tan2013}, as well as the (2x1) surface reconstruction seen by STM \cite{Wang2012}, see Fig.\,\ref{1UC_elstr} (e). Indeed, the charge transfer from the STO substrate to the FeSe layer has been detected \cite{Bang2013}. This charging fills the hole pocket of the FeSe layer and provides strong Coulomb binding between the FeSe layer and the substrate. The key component of this doping is O vacancies on the STO top layer, which are ordered along the [100] direction and strongly anchor the FeSe layer to the substrate, giving rise to a (2x1) reconstruction \cite{Bang2013}. Though, there is recent observation \cite{Ding2016} of high-$T_c$ superconductivity in 1UC FeSe films on anatase TiO2(001), with various distinct interfacial properties from STO. If confirmed, this may doubt the interfacial oxygen vacancies as the primary source for charge transfer.

The doping may also be due to charge transfer from STO impurity bands driven by work function mismatch \cite{Zhou2016}. On the other hand, the oxygen vacancies at the interface between 1UC FeSe and STO can not only provide electron doping to the interface FeSe layer, but also significantly renormalize the width of the Fe 3$d$ band near the Fermi level for the checkerboard antiferromagnetic state \cite{Chen2016}. The LDA calculations of 1UC FeSe on STO substrate \cite{Nekrasov2016} reveal the appearance of additional band of O $2p$ surface states near the Fermi level with good nesting-like matching of the hole Fe $3d$ band. Also, the 1UC-FeSe-on-STO calculations show rather small splitting of electron bands at M-point.

\begin{table}[b]
\caption{\label{tab}Difference in meV in electronic band energies in 2Fe BZ center and corner, $\Delta\varepsilon = \varepsilon(\Gamma) - \varepsilon(\textrm{M})$, for Fe $d_{xy}$, $d_{xz}$, and $d_{yz}$ bands derived from  LDA/ARPES and shown in Fig.\,\ref{Bands}.
\label{tab1}}
\begin{ruledtabular}
\begin{tabular}{lccc}
& $\Delta\varepsilon_{xy}$ & $\Delta\varepsilon_{xz}$ & $\Delta\varepsilon_{yz}$ \\
\tableline
Crystal& 560/0 & 550/20 & 470/0\\
Intercalate& 600/0 & 300/35 & 300/35\\
1UC film& 420/0 & 530/20 & 530/0\\
\end{tabular}
\end{ruledtabular}
\end{table}

Unlike the intercalates, all published ARPES spectra for 1UC FeSe show no traces of the electron pocket at the BZ center. This, however, cannot be considered as proof of its absence since all the data have been measured in-situ with helium discharge lamp, i.e. with only 21.2 eV photons \cite{Liu2012, He2013}. As a counterexample, a rare spectrum measured with 7 eV laser has been shown in \cite{Tan2013}, Fig.\,\ref{1UC_elstr} (c, bottom).

The continuous ARPES measurements during the annealing process \cite{He2013} and multi-layer film growth \cite{Tan2013} did not result in understanding of gradual evolution of the band structure from bulk to 1UC. In former, two distinct phases have been found that compete during the annealing process: the electronic structure of the phase at low doping (N phase) bears a clear resemblance to the antiferromagnetic parent compound of the Fe-based superconductors, whereas the superconducting phase (S phase) emerges with increase of doping and with suppression of the N phase \cite{He2013}. The properties of the 2UC and 1UC films are already very different. In the 2UC film the electronic structure of the interfacial FeSe layer is not affected by the surface FeSe layer, which means that the interlayer coupling and charge transfer are very weak between them \cite{Tan2013}.

Another unusual feature in ARPES spectra on 1UC on STO is the replica bands shifted at about 100 meV below the main bands. On right panel of Fig.\,\ref{1UC_elstr} (d, top) a different color scale highlights two important features: the electron band with a minimum at 60 meV below $E_F$ (labeled A), and a replica electron band (labeled A'). Panel (d, bottom) shows second derivatives in energy of the high-symmetry cuts from top row. An additional weaker replica, labeled C, can now be seen at M (right), sitting below A. At the $\Gamma$-point (left) one can see the hole band and a corresponding replica, labeled D and D', respectively \cite{Lee2014}. It has been suggested that these ``shake-off" bands appear due to presence of bosonic modes, most probably oxygen optical phonons in SrTiO$_3$ \cite{Lee2014}. Such phonons can significantly enhance the energy scale of Cooper pairing and even change the pairing symmetry \cite{Xiang2012}. These ``shake-off" bands are shown as the dotted bands on the right bottom panel of Fig.\,\ref{Bands}.

Recently, similar enhancement of superconductivity related with similar electronic structure has been found in the topmost layer in potassium-coated FeSe single crystal \cite{Ye2015}: the superconductivity emerges when the inter-pocket scattering between two electron pockets is turned on by a Lifshitz transition of Fermi surface, suggesting an underlying correlation among superconductivity, inter-pocket scattering, and nematic fluctuation in electron-doped FeSe superconductors. The results of this surface doping also confirm recent observation of the two-dome phase diagram of K-doped ultra-thin FeSe films \cite{Song2016, Miyata2015}.

\subsection{Electronic band structure summary}
\label{elstr}

\begin{table*}%[b]
\caption{\label{tab}Parameters of the hole-like $\alpha$ (in nematic state also $\nu$), $\beta$, and $\gamma$ bands and electron-like $\kappa$ band in $\Gamma$ or Z points and the hole-like $\alpha$ and $\gamma$ bands and electron-like $\delta$ and $\varepsilon$ bands in M or A points: energy of top or bottom of corresponding band $\varepsilon_0$ (eV), curvature $b$ (eV{\AA}$^2$), and associated mass $m$ (electron mass), obtained by fitting to parabola $\varepsilon(k) = \varepsilon_0 + b k^2$.
\label{tab2}}
\begin{ruledtabular}
\begin{tabular}{lcccccccccccc}
Bands && $\alpha$ &&& $\nu/\kappa/\delta$ &&& $\beta/\varepsilon$ &&& $\gamma$ &\\
& $\varepsilon_0$ & $b$ & $m$ & $\varepsilon_0$ & $b$ & $m$ & $\varepsilon_0$ & $b$ & $m$ & $\varepsilon_0$ & $b$ & $m$\\
\tableline
\textbf{Calculations}\\
FeSe $\Gamma$ LDA \cite{Maletz2014,Fedorov2016} & 0.22 & -3 & -1.3 & - & - &  & 0.17 & -12 & -0.3 & 0.11 & -2.1 & -1.8\\
FeSe M LDA \cite{Maletz2014,Fedorov2016} & -0.3 & -0.35 & -11 & -0.46 & 8.5 & 0.5 & -0.295 & 4.3 & 0.9 & -0.46 & -3.10 & -1.2\\
\\
1UC LDA $\Gamma$ \cite{Nekrasov2016} & 0.265 & -4.9 & -0.8 &  &  &  & 0.265 & -8.4 & -0.5 & -0.02 & -2.20 & -1.7\\
1UC LDA M \cite{Nekrasov2016} & -0.27 & -0.5 & -7.6 & -0.45 & 6.5 & 0.6 & -0.275 & 5.5 & 0.7 & -0.45 & -5.50 & -0.7\\
\\
1UC DMFT $\Gamma$ \cite{Nekrasov2016} & -0.03 & -6.8 & -0.6 &  &  &  & -0.03 & -6.8 & -0.6 & -0.43 & -0.80 & -4.8\\
1UC DMFT M \cite{Nekrasov2016} & -0.42 & -0.3 & -13 & -0.65 & 5.2 & 0.7 & -0.42 & 3 & 1.3 & -0.64 & -1.60 & -2.4\\
\\
\textbf{ARPES}\\
FeSe $\Gamma$ 10K \cite{Watson2015} & 0.004 & -0.93 & -4.1 & 0 & -2.5 & -1.5 & -0.05 & -5 & -0.8 & -0.05 & -0.25 & -15\\
FeSe Z 10K \cite{Watson2015} & 0.028 & -1.3 & -2.9 & 0.009 & -1 & -3.8 & -0.001 & -3 & -1.3 & -0.05 & -0.25 & -15\\
FeSe Z 120K \cite{Watson2015} & 0.02 & -1.3 & -2.9 & - & - & - & -0.001 & -3 & -1.3 & -0.035 & -0.3 & -13\\
FeSe M 37eV 10K \cite{Watson2016} & -0.002 & -0.475 & -8.0 & -0.052 & 2.1 & 1.8 & -0.008 & 4.8 & 0.8 & -0.05 & 2.10 & 1.8\\
FeSe M 37eV 96K \cite{Watson2016} & -0.01 & -0.475 & -8.0 & -0.05 & 2.3 & 1.7 & -0.018 & 4 & 1.0 & -0.04 & -0.84 & -4.5\\
\\
$({\mathrm{Tl}}_{0.58}{\mathrm{Rb}}_{0.42}){\mathrm{Fe}}_{1.72}{\mathrm{Se}}_{2}$ $\Gamma$ \cite{Mou2011} & -0.07 & -0.9 & -4.2 & -0.01 & 0.9 & 4.2 & -0.07 & -0.9 & -4.2 &  &  & \\
$({\mathrm{Tl}}_{0.58}{\mathrm{Rb}}_{0.42}){\mathrm{Fe}}_{1.72}{\mathrm{Se}}_{2}$ M \cite{Mou2011} & -0.12 & -0.3 & -13 & -0.045 & 0.52 & 7.3 &  &  &  &  &  & \\
\\
K$_{0.8}$Fe$_2$Se$_2$ $\Gamma$ \cite{Zhang2011} & -0.081 & -2.3 & -1.7 &  &  &  &  &  &  &  &  & \\
K$_{0.76}$Fe$_{1.72}$Se$_2$ $\Gamma$ \cite{Yi2015} &  &  &  &  &  &  &  &  &  & -0.055 & -0.08 & -48\\
K$_{0.8}$Fe$_2$Se$_2$ Z 31eV \cite{Zhang2011} &  &  &  & -0.018 & 1.5 & 2.5 &  &  &  &  &  & \\
K$_x$Fe$_{2-y}$Se$_2$ Z 31eV 13K \cite{Xu2012} &  &  &  & -0.03 & 6 & 0.6 &  &  &  &  &  & \\
K$_{0.8}$Fe$_2$Se$_2$ M \cite{Zhang2011} & -0.156 & -0.63 & -6.0 & -0.05 & 0.5 & 7.6 &  &  &  &  &  & \\
K$_{0.8}$Fe$_2$Se$_2$ M 26eV \cite{Zhang2011} &  &  &  &  &  &  & -0.05 & 1.1 & 3.5 &  &  & \\
\\
Rb$_x$Fe$_{2-y}$Se$_{2-z}$Te$_{z}$ $\Gamma$ \cite{Niu2015} & -0.08 & -0.8 & -4.7 & -0.008 & 2 & 1.9 &  &  &  &  &  & \\
Rb$_x$Fe$_{2-y}$Se$_{2-z}$Te$_{z}$ M \cite{Niu2015} & -0.12 & -0.25 & -15 & -0.047 & 0.52 & 7.3 &  &  &  &  &  & \\
\\
(Li$_{0.8}$Fe$_{0.2}$)OHFe$_2$Se$_2$ $\Gamma$ \cite{Niu2015} & -0.075 & -0.75 & -5.1 &  &  &  &  &  &  &  &  & \\
(Li$_{0.8}$Fe$_{0.2}$)OHFe$_2$Se$_2$ M \cite{Niu2015} & -0.11 & -0.25 & -15 & -0.043 & 0.58 & 6.6 &  &  &  &  &  & \\
\\
1UC $\Gamma$ \cite{Liu2012} & -0.07 & -1.65 & -2.3 &  &  &  &  &  &  &  &  & \\
1UC $\Gamma$ \cite{He2013} & -0.067 & -1.45 & -2.6 &  &  &  &  &  &  &  &  & \\
1UC $\Gamma$ \cite{Yi2015} &  &  &  &  &  &  &  &  &  & -0.065 & -0.2 & -19\\
1UC M \cite{Liu2012} & -0.11 & -0.6 & -6.4 &  &  &  & -0.05 & 1.1 & 3.5 &  &  & \\
1UC M \cite{He2013} & -0.113 & -0.3 & -13 &  &  &  & -0.05 & 0.9 & 4.2 &  &  & \\
 &  &  &  &  &  &  &  &  &  &  &  & \\
50ML $\Gamma$ \cite{Tan2013} & 0.02 & -1.2 & -3.2 &  &  &  & -0.016 & -2.1 & -1.8 &  &  & \\
50ML M 30K \cite{Tan2013} & -0.01 & -0.4 & -9.5 & -0.065 & 1.5 & 2.5 & -0.012 & 1.5 & 2.5 & -0.065 & -0.3 & -13\\
50ML M 115K \cite{Tan2013} & -0.012 & -0.38 & -10 & -0.049 & 1.5 & 2.5 & -0.015 & 3.4 & 1.1 & -0.049 & -0.3 & -13\\

\end{tabular}
\end{ruledtabular}
\end{table*}

Fig.\,\ref{Bands} summarizes the electronic band structure of FeSe-based compounds derived from LDA calculations (top row) and from ARPES experiments (bottom row). We tried to draw these dispersions to represent as many data as possible. Nevertheless, one can say that the most representative data sets to which we fit the dispersions on the first step, where taken from the following references: FeSe LDA \cite{Fedorov2016, Maletz2014, Watson2015} and FeSe ARPES \cite{Watson2015, Watson2016}, KFe$_2$Se$_2$ LDA \cite{Nekrasov2011} and KFe$_2$Se$_2$ ARPES \cite{Xu2012, He2013, Yi2015, Sunagawa2016}, 1UC FeSe LDA \cite{Nekrasov2016} and 1UC FeSe ARPES \cite{Yi2015}. Since we wanted to keep it clear and simple, we did not find reasonable way to show error bars here, and one need to compare it with the original spectrum to feel the level of uncertainty for each band. Still we can say that the most uncertain are the experimental $\gamma$ bands near M-point for KFe$_2$Se$_2$ and 1UC, and $\varepsilon$ band for KFe$_2$Se$_2$. The difference in meV in electronic band energies in 2Fe BZ center and corner, $\Delta\varepsilon = \varepsilon(\Gamma) - \varepsilon(\textrm{M})$, for Fe $d_{xy}$, $d_{xz}$, and $d_{yz}$ bands derived from LDA/ARPES, are given in Table\;\ref{tab1} with accuracy not better than 5 meV.

One can see that in experiment, comparing to calculations, $\Delta\varepsilon$ decreases essentially for each of the bands but to zero for $d_{xy}$ band. From Fig.\,\ref{Bands} one can see that in terms of electron hopping, the $d_{xy}$ and $d_{xz}$ bands along the $\Gamma$M direction can be well approximated by two nearest neighbors,
$$\varepsilon(k) = \varepsilon_0 + t_1 \cos(ka) + t_2 \cos(2ka),$$
where $a$ is the Fe-Fe distance, $t_1$, and $t_2$ are the hopping integrals. Within this oversimplified approximation, $\Delta\varepsilon = t_1 = 0$ would mean that the hopping between the nearest neighbors is blocked completely. Naturally, the peculiar for Fe-SC's spin or orbital orderings, even in form of spin/orbital-fluctuations, should suppress the near-neighbor hopping, but in order to understand its complete suppression, more sophisticated model should be elaborated.

As for proximity to Lifshitz transitions, only FeSe, and to a greater extent Fe(Se,Te), may belong to $\Gamma_h$-class of Fe-SC's, similarly to LiFeAs and Ba(Fe,Co)$_2$As$_2$, where superconductivity could be enhanced by a shape resonance \cite{Bianconi2013} on small 3D FS formed by the inner hole ($\beta$) band. One may speculate that similar $T_c$ enhancement one can get on $\kappa$ band in intercalates and on some yet hidden bands in 1UC films, but this issue certainly requires further investigation.

In Table\;\ref{tab2} we give the parameters of different bands obtained by fitting the experimental dispersions around $\Gamma$, Z, or M-points with parabolic dispersions $\varepsilon(k) = \varepsilon_0 + b k^2$: $\varepsilon_0$ (eV), curvature $b$ (eV{\AA}$^2$), and associated mass $m$ (electron mass). In the first column together with the name of the compound we show also the type of the data (LDA, DMFT, or ARPES), and, in case of ARPES, the photon energy and temperature of the sample. The values of $\varepsilon_0$ were used to evaluate the proximity of certain bands to Lifshitz transition, but does it make any sense to discuss the band curvature $b$ or the mass $m \propto 1/b$? These values are often used to evaluate renormalization: $1 + \lambda = m_{\textrm{ARPES}}/m_{\textrm{LDA}}$. Since all the bands in Fe-SC's are not only squeezed in energy but also deformed by the ``red-blue shift", as shown schematically in Fig.\,\ref{FeSe_elstr} (b), one needs to either include such a shift in renormalization or take it into account as external effect. The former could be described as a hopping selective renormalization, meaning that electronic interaction differently affects different hopping integrals. In the case of the two nearest neighbor hopping model, the band associated masses in $\Gamma$ and M points would be different: $m_\Gamma \propto (-t_1 - 4 t_2)^{-1}$, $m_\textrm{M} \propto (t_1 - 4 t_2)^{-1}$. It is indeed the case for FeSe, as one can see from Table\;\ref{tab2}. For example, for FeSe single crystals at 10 K for $\gamma$ band $m_{\textrm{ARPES}}/m_{\textrm{LDA}} \approx 8$ in $\Gamma$ but -1.5 in M-point at 10 K and about 4 at 96 K. So, one may conclude that evaluation of the renormalization from experiment only makes sense if accompanied by a model of interaction.

\section{Nematicity and temperature evolution}
\label{Nematicity}

\begin{figure*}
\begin{center}
\includegraphics[width=1\textwidth]{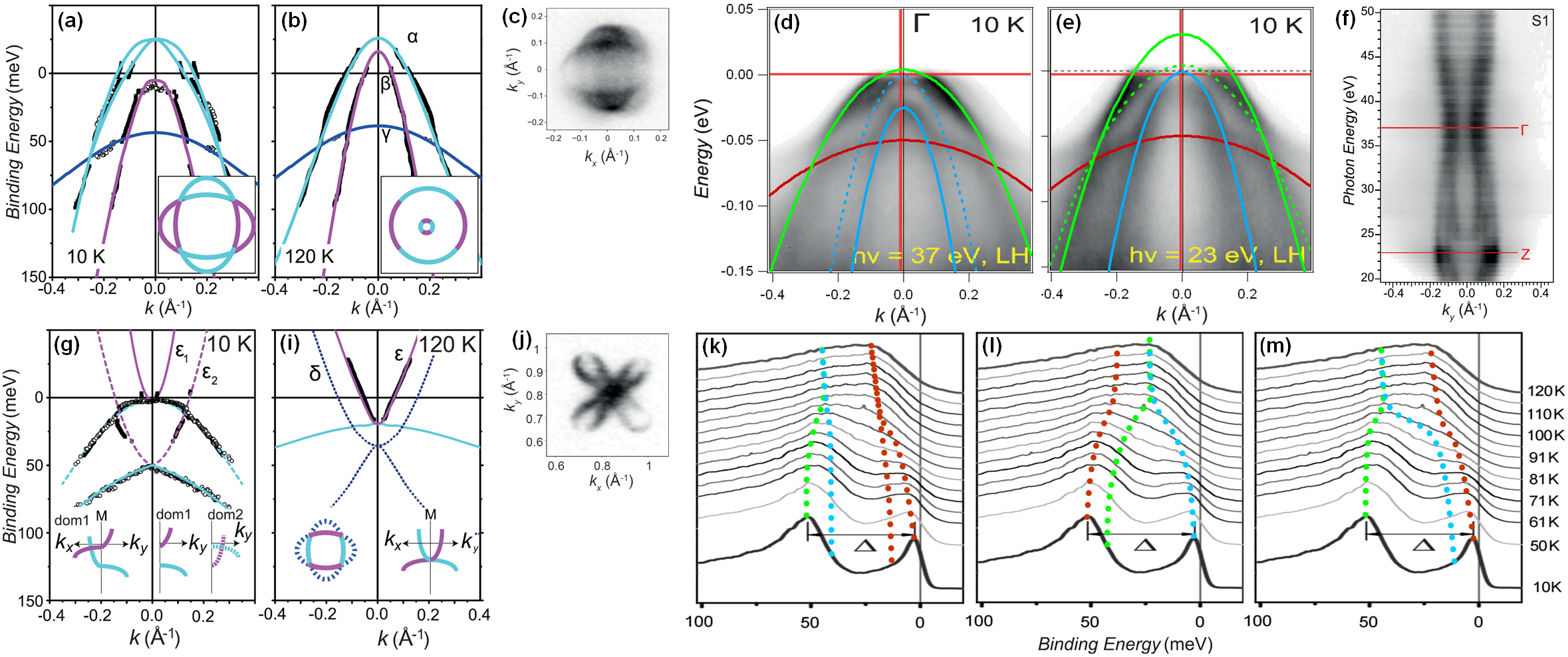}
\caption{Nematicity in FeSe crystals around $\Gamma$Z axis (a-f) and around M-point (g-m) \protect\ignorecitefornumbering{\cite{Watson2015}}.
Experimental dispersions in $\Gamma$M direction below (a, g) and above (b, i) nematic transition around $\Gamma$ (a, b) and M-point (g, i).
Experimental FS's around $\Gamma$ (c) and M-point (j), as well as, the photon energy dependence of ARPES intensity around BZ center that represents the experimental FS in ($k_x$, 0, $k_z$) plane (f).
Parabolic fits to the experimental hole bands at $\Gamma$ (d) and Z (e) shows (1) that the two outer bands cannot meet at their tops and (2) that the middle (dashed blue) band near $\Gamma$-point (d) coincides with the inner band (solid blue) from near Z-point, and the middle band (dashed green) near Z-point (e) coincides with the outer (solid green) band from the $\Gamma$-point.
Three possible scenario of band splitting at $T_s$ over the temperature dependence of EDC taken at M-point (k-m).
The color curves in panels (d, e, k-m) are put over the experimental data taken from \protect\ignorecitefornumbering{\cite{Watson2015}}.
\label{fignemat}}
\end{center}
\end{figure*}

\textbf{Nematicity.} To address the nematicity issue we should go back to single crystals. Fig.\,\ref{Res} contains a number of examples how the nematic transition is seen in electro-transport measurements. In diffraction experiments this transition appears as spontaneous breaking of the symmetry between the $x$ and $y$ directions in the Fe-plane but it has been said that its effect on electronic properties is much larger than expected based on the structural distortion observed \cite{Tanatar2010, Chu2010}. This was the main argument to relate this structural transition with intrinsic electronic instability and call ``nematic".

Five years later, when the quality of FeSe crystals had been considerably improved \cite{Chareev2013, Watson2015, Roessler2016}, the electronic structure in nematic state has been revealed and studied in a number of ARPES papers \cite{Shimojima2014, Nakayama2014, Watson2015}. Fig.\,\ref{FeSe_elstr} (d-h) and Fig.\,\ref{fignemat} show the results of the most accurate ARPES study of the nematic phase as for the middle of 2015 \cite{Watson2015}. The effects of nematicity have been observed in both $\Gamma$(Z)- and M-regions of BZ. In $\Gamma$(Z)-region the circular FS, see inset in Fig.\,\ref{fignemat} (b), splits into two ellipses below $T_s$, Fig.\,\ref{fignemat} (a, c). Each ellipse is attributed to a set of orthogonal domains, so, the corresponding dispersions should meet on $\Gamma$Z-axis of BZ. Since the electronic structure in the corners of the orthorhombically distorted BZ may be different, the large splitting of about 50 meV seen in M-point, Fig.\,\ref{fignemat} (g), has been considered as huge effect of nematicity \cite{Shimojima2014, Nakayama2014, Watson2015} and supported by many other ARPES studies \cite{Zhang2015, YZhang2015, Fanfarillo2016}.

However, even in the original set of spectra from M-point measured at different temperature \cite{Watson2015}, one can see that the final splitting is present in all curves above $T_s = 90 \textrm{K}$: Fig.\,\ref{fignemat} (k-m) show this set of spectra with 3 possible scenario of peaks splitting. The authors of Ref.\,\onlinecite{Watson2015} have noticed the final splitting above $T_s$ and have resolved it in recent experiment \cite{Watson2016}, but they suggest another scenario, where now no additional splitting appears but the separation between the $d_{xy}$ and $d_{xz}/d_{yz}$ bands increases below $T_s$ due to ``unidirectional nematic bond ordering". We think that the most reasonable is the scenario shown in Fig.\,\ref{fignemat} (k) when both $d_{xy}$ and $d_{xz}/d_{yz}$ bands split below $T_s$ of about 10 meV. This scenario is supported by recent data \cite{Fedorov2016}, where the nematicity driven splitting is estimated as 15 meV and is also in line with previous theoretical arguments \cite{Fernandes2014}. Nevertheless, we think that there is still space for all the scenarios. The point is that the lineshape of each of the two peaks below $T_s$ are evidently not a shape of the single band spectral function, but it should not be so due to strong $k_z$ dependence of the lower band (see Fig.\,\ref{FeSe_elstr}a) and certain $k_z$ integration due to very finite electron escape depth \cite{Kordyuk2014}.

There are also different opinions about FS splitting in the BZ center. The results of Ref.\,\onlinecite{Fedorov2016} support the nematic domains scenario \cite{Watson2015}, while other authors, e.g. \cite{YZhang2015}, believe that the splittings at the $\Gamma/$Z and M points are controlled by different order parameters. In Fig.\,\ref{fignemat} (d,e) we show the parabolic fits to the experimental hole bands at $\Gamma$ (d) and Z (e). Based on these fits, we want to note that it is unlikely that the two outer bands meet at their tops, as it would be required by the nematic domain scenario. This is especially clear in Z-point. Also, the middle (dashed blue) band near $\Gamma$-point  coincides with the inner band (solid blue) from near Z-point, and the middle band (dashed green) near Z-point (e) coincides with the outer (solid green) band from the $\Gamma$-point. This may be a result of doubling of the unit cell or similar suppression of the near neighbor hopping in $k_z$ direction. Interestingly, the photon energy dependence of the experimentally measured FS shows a twice smaller period for the inner FS corrugation. The importance of the out-of-plane interaction for nematic ordering follows also from the fact that nematicity is observed in single crystals and in multi layer films \cite{Zhang2015} but neither in 1UC films nor in intercalates.

\textbf{Temperature evolution.} As we have shown above, the electronic band structure of bulk FeSe compounds changes dramatically with temperature across the nematic transition. Interestingly, the band structure continues to evolve far above \cite{AbdelHafiez2016} (and far below) $T_s$. Also we have discussed that the real band structure of Fe-SC's, measured usually at low temperature, in addition to strong renormalization, is also deformed by a ``red-blue shift" with respect to first-principle calculations, as shown schematically in Fig.\,\ref{FeSe_elstr} (b). Such shifts are observed for all FeSC's, in particular for BFCA \cite{Brouet2013}, BKFA, and LiFeAs \cite{Kordyuk2012, Kordyuk2013} and can be described in terms of hopping selective renormalization or as Pomeranchuk instability of the Fermi surface \cite{Zhai2009, Massat2016}. On the other hand, such a shift can be natural consequence of the strong particle-hole asymmetry in multiband nearly compensated metals \cite{Ortenzi2009, Benfatto2011, Fanfarillo2016}.

In any case, it is tempting to suppose that at hight enough temperature, the band structure, or at least the FS topology, coincides with the result of LDA band structure calculations but an interaction like the hopping selective renormalization develops with lowering temperature and with increase of the strength of fluctuations of certain order. While the temperature dependent ARPES is rather complicated \cite{Kordyuk2014}, one may consider the temperature dependent Hall measurements as a complementary tool \cite{Evtushinsky2008} that is especially sensitive when the FS goes through the topological Lifshitz transitions, such as from electron like barrels to mixed electron-hole like propellers in BKFA \cite{Evtushinsky2011, Zabolotnyy2009}. Then, careful temperature dependent ARPES measurements would be a key tool to identify the microscopic interaction that is responsible for both the temperature evolution of the band structure in Fe-SC's and the nematic instability.

\section{Conclusions}
\label{concl}

In this review, we have analyzed the published results on electronic structure of FeSe-based superconductors: isovalently doped crystals, intercalates, and single layer films. We have summarized the results of first-principle calculations and ARPES experiments in Fig.\,\ref{Bands} and Tables I and II. The experimental band structure, with respect to calculations, is renormalized and selectively shifted (``red-blue shift") such as the Fe-Fe nearest neighbor hopping is suppressed essentially for the Fe 3$d_{xz}$ conducting band and suppressed completely for the $d_{xy}$ band for all the FeSe compounds here considered. This suggests a crucial role of spin or orbital fluctuations in formation of the normal state electronic structure as a ground state for superconductivity. Temperature dependent ARPES measurements through the nematic transition and to the highest temperatures will be a key tool to identify the microscopic interaction responsible for those shifts.

The nematicity is certainly a result of intrinsic electronic instability but the details of its effect on electronic structure is still controversial. In particular, the role of three-dimensionality is not studied at all.

As for proximity to Lifshitz transitions, only FeSe, and to a greater extent Fe(Se,Te), may belong to $\Gamma_h$-class of Fe-SC's (see Fig.\,\ref{UPhD}), similarly to LiFeAs and Ba(Fe,Co)$_2$As$_2$, where superconductivity could be enhanced by a shape resonance on small 3D FS formed by the inner hole band. One may speculate that similar $T_c$ enhancement one can get on small electron FS in intercalated samples and on some yet hidden bands in 1UC films, but this issue certainly requires further investigation.

\begin{acknowledgements}
We acknowledge discussions with M. Abdel-Hafiez, V. V. Bezguba, A. Bianconi, S. V. Borisenko, V. Brouet, D. A. Chareev, A. V. Chubukov, A. I. Coldea, I. Eremin, D. V. Evtushinsky, D. S. Inosov, T. K. Kim, M. M. Korshunov, I. V. Morozov, S. R\"o{\ss}ler, M. V. Sadovskii, J. Spa{\l}ek, A. N. Vasiliev, A. N. Yaresko, and V. B. Zabolotnyy.
\end{acknowledgements}

\bibliographystyle{PRL}
\bibliography{FeSe_Rev}

%\onecolumngrid
\end{document}